\documentclass[10pt]{article}
\usepackage{amssymb}  
\usepackage{amsfonts}
\usepackage{graphicx}
\usepackage{amsmath}
\usepackage{authblk}
\usepackage{bm}
 
\newtheorem{prop}{Proposition}[section] 
 
\newtheorem{rem}{Remark}[section]

\newtheorem{propr}{Property}[section]
\def\x2dot{\mathop{x}\limits}
\def\y2dot{\mathop{y}\limits}
\def\bfy2dot{\mathop{\bf y}\limits}
\def\z2dot{\mathop{z}\limits}
\def\min{\mathop{min}\limits}
\def\max{\mathop{max}\limits}
\def\csi2dot{\mathop{\xi}\limits}
\def\et2dot{\mathop{\eta}\limits}
\def\bet2dot{\mathop{\beta}\limits}
\def\t2dot{\mathop{\theta}\limits}
\def\s2dot{\mathop{\sigma}\limits}
\def\d2dot{\mathop{\delta}\limits}
\def\l2dot{\mathop{\lambda}\limits}
\def\ps2dot{\mathop{\psi}\limits}
\def\tet2dot{\mathop{\theta}\limits}
\def\bfy2dot{\mathop{\bf y}\limits}
\def\bfq2dot{\mathop{\bf q}\limits}
\def\w2{\mathop{W}\limits}
\def\xgrande2dot{\mathop{\bf X}\limits}
\def\p02dot{\mathop{P}\limits}
\def\a2dot{\mathop{A}\limits}

\title{Synchronization of two coupled pendula in absence of escapement}
\author{Federico Talamucci}
\affil{DIMAI, Dipartimento di Matematica e Informatica ``Ulisse Dini'',\\
Universit\`a degli Studi di Firenze, Italy\\
tel.~+39 055 2751432, fax +39 055 2751452
\\
e-mail: federico.talamucci@math.unifi.it}
\date{}

\begin{document}
\bibliographystyle{plain}

\setcounter{equation}{0}
\setcounter{ese}{0}
\setcounter{eserc}{0}
\setcounter{teo}{0}               
\setcounter{corol}{0}
\setcounter{propr}{0}

\maketitle

\vspace{.5truecm}


\vspace{.5truecm}


\vspace{.5truecm}

\noindent
{\bf Abstract}. A model of two oscillating pendula placed on a mobile support is studied. 
Once an overall scheme of equations, under general assumptions, is formulated via the Lagrangian equations of motion, the specific case of absence of escapement is examined. 

\noindent
The mechanical models consists of two coupled pendula both oscillating on a moving board 
attached to a spring.

\noindent
The final result performs a selection among the peculiar parameters of the physical process (lenghts, ratio of masses, friction and damping coefficients, stiffness of the spring) 
which provide a tendency to synchronization.

%%%%%%%%%%%%%%%%%%%%%%%%%%%%%%%%%%%%%%%%%%%%%%%%%%%%%%%%%%%%%%%%%%%%%%%%%%%%%%%%%%%%%%%%%%

\section{Introduction}

\noindent
Systems of coupled oscillations are largely studied on account of their  wide 
possibility of application in many significant branches (mechanics, medical and byological
sciences, ...).
The corresponding mathematical problem is in no way easy to handle when all the effects are overlapped: here, we propose a basic situation which will be discussed from the mathematical point of view.

\noindent
The main question we deal with is the feasibility of in--phase or antiphase synchronization when no external forces (escapement) forcing the free oscillation are contemplated.

\noindent
We mainly take care of the mathematical path drawn by the equations of motion, 
aiming at developing the analytical scheme, even if in a simplified situation.
We first formulate the mathematical model by allowing very general features of the 
mechanical phenomenon, admitting different sizes of masses, lenghts of the pendula and including escapement conditioning the oscillations.
This will supply a ground in order to make a brief comparison with some significant models proposed in literature.

\noindent
Rather than obtaining information directly via a numerical simulation
approach, we rather aim for via an analytical method of locating the eigenvalues 
linked with the damping of the system.
The advantage is the prospect of recognizing some ranges of the parameters entering the phenomenon which predispose the sytem to synchronization.

\noindent
On the other hand, some expected results are confirmed, as the unattainability of the in--phase synchronization in absence of escapement.

\noindent
The first step, introduced in the next Section, is the mathematical formulation of the model, which will be achieved by the Lagrangian's method of deducing the equations of motion.

\section{The mathematical model}

\noindent
The system we are going to study can be realized either by device $I$ or device $II$ shown in Figure 1: as for $I$, the apparatus consists of two pendula whose pivots 
(points $A$ and $B$) are fasten on a horizontal and homogeneous beam with mass $m_0$ 
and $P_0$ as centre of mass. The beam is placed on a pair of rollers of radius $R$. 
The massive bobs (of mass $m_1$ and $m_2$) are suspended at the extremities $P_1$ and $P_2$ of the massless rods (of lenghts $\ell_1<R$ and $\ell_2<R$) and can oscillate on the vertical plane containing the beam. The distances $d_1$, $d_2$ of $P_1$ and $P_2$ from $P_0$ are larger than $\ell_1$ and $\ell_2$, in order to avoid hits. A spring of stiffness $k$ and whose mass and lenght at rest are negligible, is attached at one of the extremities of the beam.

\begin{figure}[htbp]
\includegraphics[scale=.39]{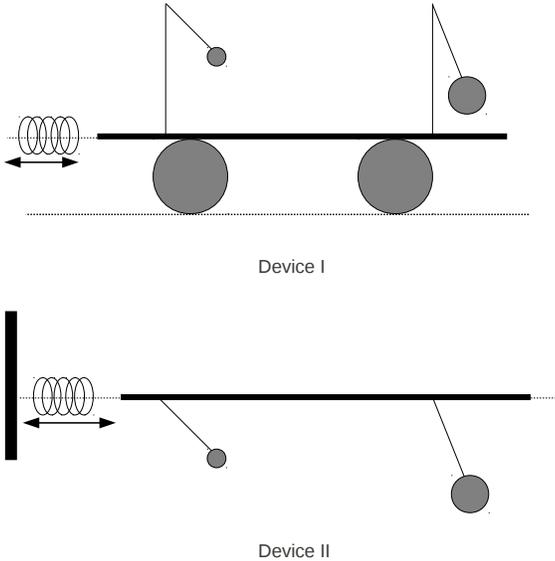}
\caption{The board supporting the two pendula leans against a couple of rollers (device $I$) or moves back and forth on a horizontal support (device $II$). In both cases, a spring is attached to one of the exteremities of the board.}
\end{figure}

\noindent
In device $II$ the pendula are suspended by means of a T--shaped support, with negligible mass and fixed on the beam which oscillates at a lower height. 

\noindent
If the physical quantities $m_0$, $m_1$, $m_2$, $\ell_1$, $\ell_2$ are the same in $I$ and in $II$, it is immediate to realize that the mathematical problem is identical (actually the lenghts $h$, $R$, $d_1$ and $d_2$ do not enter the equations of motion).
On the other hand, not even from the dynamical point of view the two systems are different: actually, active forces are the same and, assuming that the rolling friction in apparatus $I$ is proportional to ${\dot \Psi}$, with $\Psi$ rotation angle of the roll, then it is proportional to ${\dot x}$, with $x$ abscissa of $P_0$ (namely $x-2R\Psi=constant$), just like in apparatus $II$.
 
\subsection{The equations of motion}

\noindent
In this first part the equations of motion are achieved by using a Lagrangian approach. 
The cartesian coordinate system is fixed in order that the mechanism is contained 
in the vertical plane $y=0$ and the beam swings along the $x$--axis, the $z$--axis is 
upward--vertically directed, the origin $O$ corresponds to the fixed extremity of the spring.

\noindent
We choose as lagrangian coordinates ${\bf q}=(x, \theta_1, \theta_2)$, where $\theta_1$ and $\theta_2$ are the amplitudes of oscillation with respect to the downward--vertical direction and $x$ is the abscissa of $P_0$: the representative vector of the discrete system $(P_0, P_1, P_2)$ in terms of them is, for device $I$, 
$$
{\bf X}({\bf q})= (x,0,2R,x-d_1+\ell_1 \sin \theta_1, 0 , 2R-\ell_1 \cos \theta_1, 
x+d_2+\ell_2 \sin \theta_2, 0 , 2R-\ell_2 \cos \theta_2)
$$ 
and $0$ at the third position, $h$ replacing $2R$ for device $II$.
In both cases, the Lagrangian function ${\cal L}=T+U$, where $U$ is the potential of the elastic force and of gravity, is
\begin{equation}
\label{lagr}
{\cal L}({\bf q}, {\dot {\bf q}})=\dfrac{1}{2} m{\dot x}^2 +\sum\limits_{j=1}^2 \left( \dfrac{1}{2}m_j \ell_j^2{\dot \theta}_j^2+
m_j \ell_j\left( {\dot x}{\dot \theta}_j+g\right) \cos \theta_j \right)-\dfrac{1}{2}kx^2
\end{equation}
where $m=\sum\limits_{j=0}^2 m_j$ is the total mass.

\noindent
As for the friction forces, 
if the damping is formulated as ${\bm \Phi}_{P_i}=-\beta_i{\dot P}_i$, $i=0,1,2$, the lagrangian components are
\begin{equation}
\label{damping1}
{\bm \Phi}^{({\bf q})}=
\left(-\beta {\dot x} -\sum\limits_{j=1}^2 \beta_j \ell_j {\dot \theta}_j \cos \theta_j, -\beta_1 \ell_1\left( {\dot x}\cos \theta_1 + \ell_1 {\dot \theta}_1\right),  
-\beta_2 \ell_2\left( {\dot x}\cos \theta_2 + \ell_2 {\dot \theta}_2\right)
\right)
%=D({\dot {\bf q}})
\end{equation}
with $\beta=\sum\limits_{j=0}^2\beta_j$.  
If, on the contrary, one assumes that the pendula run into damping only along the rotational direction  ${\bf e}_{\theta_i}=(\cos \theta_i, 0, \sin \theta_i)$, $i=1,2$, then the generalized friction force reduces to
\begin{equation}
\label{damping2}
{\bm \Phi}^{({\bf q})}=
\left(-\beta_0 {\dot x} -\sum\limits_{j=1}^2 \beta_j \ell_j {\dot \theta}_j \cos \theta_j,
 -\beta_1 \ell_1^2 {\dot \theta}_1, -\beta_2 \ell_2^2 {\dot \theta}_2,\right).
%={\widehat D}({\bf q})
\end{equation}
%where ${\widehat D}$ is $D$, except for the first column, which is now $(-\beta_0, 0,0)^T$.

\noindent
The mechanism of escapement can be modelled by introducing a moment in the direction ${\bf j}$ (i.~e.~orthogonal to the plane of the system), exerting a force $f_i(P_i, {\dot P}_i, t){\bf e}_{\theta_i}$ on each $P_i$, $i=1,2$.
The corresponding generalized force is 
\begin{equation}
\label{escapement}
{\bf F}^{({\bf q})}=
\left(f_1 \ell_1 \cos \theta_1+f_2\ell_2 \cos \theta_2, \ell_1 f_1, \ell_2 f_2\right).
\end{equation}
Assuming (\ref{damping1}), the equations of motion $\dfrac{d}{dt}\left( \nabla_{\dot {\bf q}}{\cal L}\right)-
\nabla_{\bf q}{\cal L}={\bf F}^{({\bf q})}+{\bm \Phi}^{({\bf q})}$ are
\begin{equation}
\label{eqlagr}
\left\{
\begin{array}{l}
m\x2dot^{..} +\sum\limits_{j=1}^2 \ell_jm_j\left( {\t2dot^{..}}_j \cos \theta_j - 
{\dot \theta}_j^2\sin \theta_j\right)+kx=-\beta {\dot x} +\sum\limits_{j=1}^2\ell_j
\left( f_j-\beta_j {\dot \theta}_j\right) \cos \theta_j, \\
m_i\ell_i \left(\ell_i {\t2dot^{..}}_i + \x2dot^{..}\cos \theta_i +g\sin \theta_i \right)=
\ell_i \left( f_i - \beta_i \left({\dot x}\cos \theta_i+\ell_i {\dot \theta}_i\right) \right)
\qquad i=1,2
\end{array}
\right.
\end{equation}

\subsection{Comparison with some models}

\noindent 
In reviewing very briefly the mathematical formulation of some models existing in literature, we have the specific intention of remarking that
\begin{description}
\item[$(1)$]
if (\ref{damping2}) is accepted to hold, then $\beta_0$ replaces $\beta$ in the first equation and the terms $-\ell_i \beta_i{\dot x}\cos \theta_i$, $i=1,2$, have to be omitted in the second and third equation. However, the term
$-\sum\limits_{j=1}^2\ell_j \beta_j{\dot \theta}_j\cos \theta_j$ of the first equation is 
present anyway;
\item[$(2)$] even if the escapement mechanism operates on the pendula vie the force (\ref{escapement}), terms containing $f_1$ and $f_2$ are present in any case in the
first equation in (\ref{eqlagr}) concerning $x$ and they affect the motion of the beam.
\end{description} 

\noindent
In \cite{czo}, where the apparatus $I$ is tested, the escapement is formulated by means of a step function, depending on the amplitude of a threshold angle. In \cite{ben} the apparatus $II$ is subject to the inversion of direction of the angular velocity (escapement mechanism) at a critical value of the angle.
Also in \cite{oud} the mathematical problem is formulated for two driven pendula
(although the description of the experimental setup refers to a couple of metronomes), but the function (\ref{escapement}) is expressed via a continous function. 

\noindent
The cited models are undoubtedly significant and useful for the exhibited experimental and numerical results: nevertheless, we remark the lack of the terms in (\ref{eqlagr}), first equation, pointed out in  $(1)$, $(2)$ just above. This aspect should not be unimportant, mainly when analytical results are pursued, as in our investigation.

\noindent
An experimental device which differs from $I$ and $II$ described at the beginning of the Section consists in placing two masses at each of the pivotal points and let them oscillate 
horizontally, by means of a spring connecting the two points (hence the beam is removed): 
such as interaction mechanism is studied in \cite{dil} and the analytical problem is 
essentially the same as (\ref{eqlagr}), since the centre of mass of the attachement points 
solves the first of such system, with $k=0$. 

\noindent
Finally, in the system studied in \cite{fra} and \cite{kum} the two pendula are coupled by a spring connecting some intermediate points ($Q_1$ and $Q_2$) of the two sticks supporting the weights (Kumamoto coupled pendula).
The spring--coupled pendula are proposed also as a basic model for the neutrino oscillation.

\noindent
Calling $d$ the constant distance between the pivotal points and presuming that the lenght at rest of the 
spring is $d$, we write the Lagrangian function as $\dfrac{1}{2} m_1 \ell_1^2 {\dot \theta}_1^2+\dfrac{1}{2}m_2 {\dot \theta}_2^2 +m_1 g \ell_1 \cos \theta_1 +m_2 \ell_2 g \cos \theta_2 +\dfrac{1}{2}k\left( |Q_1(\theta_1)-Q_2(\theta_2)|-d\right)^2$ where
$$
|Q_1-Q_2|=\left(d^2+\ell_{0,1}^2+\ell_{0,2}^2+2d(\ell_{0,1}\sin \theta_1 -\ell_{0,2}\sin \theta_2)-2\ell_{0,1}\ell_{0,2} \cos (\theta_1 - \theta_2)\right)^{1/2}
$$
being $\ell_{0,1}$, $\ell_{0,2}$ the distances between the pivots and the intermediate points.

\noindent
The problem is here two--dimensional and the corresponding equations of motion for the two angles, by assuming (\ref{damping2}) and $\ell_1 f_1 =u(t)$, $f_2=0$ (external torque acting only on one pendulum, see \cite{kum}) are
$$
\begin{array}{l}
m_1 \ell^2_1 {\t2dot^{..}}_1 + m_1 g \ell_1 \sin \theta_1 
+\ell_{1,0} k [d\cos \theta_1 +\ell_{2,0} \sin (\theta_1 - \theta_2)]
\left(1-\dfrac{d}{|Q_1-Q_2|}\right)= 
u(t)- \ell_1^2\beta_1 {\dot \theta}_1^2, \\
m_2 \ell^2_2 {\t2dot^{..}}_2 + m_2 g \ell_2 \sin \theta_2
-\ell_{2,0} k[d\cos \theta_2 +\ell_{1,0} \sin (\theta_1 - \theta_2)]
\left(1-\dfrac{d}{|Q_1-Q_2|}\right)=
- \ell_2^2\beta_2 {\dot \theta}_2^2
\end{array}
$$
which slightly differ from the dynamics formulated in \cite{kum},
where $\ell_{0,1}=\ell_{0,2}$ and $d$ is neglected.

\subsection{The mathematical problem for $\sigma$ and $\delta$}

\noindent
The main purpose is to investigate the existence of solutions of system (\ref{eqlagr}) 
such that one of the the quantities 
\begin{equation}
\label{sigmadelta}
\sigma= \theta_1 + \theta_2, \quad 
\delta= \theta_1 -  \theta_2
\end{equation}
tends to zero for $t\rightarrow +\infty$. If $\delta\rightarrow 0$ [respectively $\sigma\rightarrow 0$], then the system will proceed to in--phase synchronization [resp.~antiphase synchronization].

\noindent
Moving toward the more expressive coordinates (\ref{sigmadelta}) and setting 
${\bf y}=\left( \begin{array}{c} x \\ \sigma \\ \delta \end{array}\right)$, one has ${\bf y}=L{\bf q}$, where $L=$ {\footnotesize $\left(\begin{array}{ccc} 1 & 0 & 0\\
0 & 1 & 1  \\ 0 & 1 & -1 \end{array}\right)$} is the linear change of coordinates. Thus,  
writing again the equations of motion by taking account of
$
\dfrac{d}{dt}\left( 
\nabla_{\dot {\bf y}}{\widetilde {\cal L}}\right)-
\nabla_{\bf y}{\widetilde {\cal L}}
=
L^{-1} \left( 
\dfrac{d}{dt}\left( \nabla_{\dot {\bf q}}{\cal L}\right)-
\nabla_{\bf q}{\cal L}\right)$, 
${\bf F}^{({\bf y})}+{\bm \Phi}^{({\bf y})}=L^{-1}\left(
{\bf F}^{({\bf q})}+{\bm \Phi}^{({\bf q})}\right)$
with ${\widetilde {\cal L}}({\bf y}, {\dot {\bf y}})={\cal L}(L^{-1}{\bf y}, L^{-1}{\dot {\bf y}})$, $L^{-1}={\footnotesize \left(\begin{array}{ccc} 1 & 0 & 0\\
0 & 1/2 & 1/2  \\ 0 & 1/2 & -1/2 \end{array}\right)}$, we attain

\begin{equation}
\label{eqlagrsd}
\left\{ 
\begin{array}{l}
m\x2dot^{..} +
\Psi_1(B_m^+, -B_m^-)\s2dot^{..}+\Psi_1 (B_m^-,-B_m^+)\d2dot^{..}=
\dfrac{1}{2} \Psi_2 (B_m^+, B_m^-)({\dot \sigma}^2+{\dot \delta}^2)+
\Psi_2(B_m^-, B_m^+){\dot \sigma}{\dot \delta}-Kx+\\
+\Psi_1 (f_1+f_2, f_2-f_1)-\beta {\dot x} -\Psi_1 (B_\beta^+, -B_\beta^-){\dot \sigma} 
- \Psi_1 (B_\beta^-, -B_\beta^+){\dot \delta},
\\
\\
\Psi_1 (B_m^+, -B_m^-)\x2dot^{..}+
A_m^+\s2dot^{..}+A_m^-\d2dot^{..} = -g \Psi_2 (B_m^+, B_m^-)
+f^+ - \Psi_1 (B_\beta^+,-B_\beta^-){\dot x}
-A_\beta^+{\dot \sigma} -A_\beta^-{\dot \delta},
\\
\\
\Psi_1 (B_m^-,-B_m^+)\x2dot^{..}+A_m^-\s2dot^{..}+A_m^+\d2dot^{..} = 
-g \Psi_2 ( B_m^-\, B_m^+)
+f^- -\Psi_1 (B_\beta^-, -B_\beta^+){\dot x}
-A_\beta^-{\dot \sigma} -A_\beta^+{\dot \delta} 
\end{array}
\right.
\end{equation}
where 
\begin{equation}
\label{psi12}
\begin{array}{l}
\Psi_1 (\sigma, \delta; C_1, C_2)=C_1 \cos (\sigma/2) \cos (\delta/2)+C_2 
\sin  (\sigma/2) \sin  (\delta/2), \\
\Psi_2 (\sigma, \delta; C_1, C_2)=C_1 \sin  (\sigma/2) \cos (\delta/2)+C_2 \cos  (\sigma/2) \sin (\delta/2), \qquad  C_1, C_2 \in {\Bbb R}
\end{array}
\end{equation}
and 
\begin{equation}
\label{pm}
\begin{array}{llll}
A_m^{\pm}=\dfrac{1}{4}(m_1\ell_1^2\pm m_2 \ell_2^2), & 
A_\beta^{\pm}=\dfrac{1}{4} (\beta_1 \ell_1^2\pm \beta_2 \ell_2^2), &
B_m^{\pm}=\dfrac{1}{2} (m_1 \ell_1 \pm m_2 \ell_2), &
B_\beta^{\pm}=\dfrac{1}{2} (\beta_1 \ell_1 \pm \beta_2 \ell_2),\\
f^{\pm}=\dfrac{1}{2} (\ell_1 f_1 \pm \ell_2 f_2).
\end{array}
\end{equation}

\noindent
It is worth noticing that
\begin{description}

\item[$(i)$] if (\ref{damping2}) holds instead of (\ref{damping1}), then $\beta_0$ replaces $\beta$ after the equals sign in the first equation and each last term of second and third equation (containing ${\dot x}$) has to be removed.

\item[$(ii)$]
In case of identical pendula ($m_1=m_2$, $\ell_1=\ell_2$) all the quantities in (\ref{pm}) with superscript $-$ vanish, so that each term in (\ref{eqlagrsd}) containing them 
must be cancelled.

\item[$(iii)$]
Likewise, if also $f_1=f_2$ additional simplifications are evident.

\item[$(iv)$] 
Solving (\ref{eqlagrsd}) explicitly with respect to the second order derivatives one gets 
\begin{equation}
\label{eqlagrsdfn}
\left\{
\begin{array}{l}
{\mit \Theta} \x2dot^{..} =
-kx +\dfrac{1}{2} g \left[ (m_1+m_2)\sin \sigma \cos \delta + (m_1 - m_2) \sin \delta 
\cos \sigma \right] +
\dfrac{1}{2} 
\left[
\Psi_2 (B_m^+, B_m^-)({\dot \sigma}^2+{\dot \delta}^2)+\right.\\
\left.
+2\Psi_2 (B_m^-, B_m^+){\dot \sigma}{\dot \delta} 
\right]-\left[\beta_0 
+\dfrac{1}{2}(\beta_1+\beta_2)-\dfrac{1}{2}\left( (\beta_1+\beta_2) \cos\sigma\cos \delta+(\beta_2-\beta_1) \sin \sigma \sin \delta\right)\right]{\dot x}, 
\\
\\
\Theta \s2dot^{..}=
-2g\left[m\Psi_2 (\ell^+,-\ell^-)-
\dfrac{1}{\ell_1\ell_2}\Psi_1 (-B_m^-, B_m^+)\sin \delta\right]
+2k\Psi_1(\ell^+, \ell^-)x-\\
- \Psi_1 (\ell^+,\ell^-)\left[
\Psi_2 (B_m^+, B_m^-)({\dot \sigma}^2+{\dot \delta}^2)
+2\Psi_2 (B_m^-, B_m^+){\dot \sigma \dot \delta} 
\right]+\\
+\Theta \left(\dfrac{f_1}{m_1\ell_1}+\dfrac{f_2}{m_2\ell_2}\right)+
\left[2\beta \Psi_1 (\ell^+,\ell^-)-m\Psi_1 
\left(\dfrac{\beta_1}{m_1 \ell_1}+\dfrac{\beta_2}{m_2 \ell_2}, 
\dfrac{\beta_2}{m_2 \ell_2} -\dfrac{\beta_1}{m_1 \ell_1}
\right)
-\right.\\
\left.
-\dfrac{2\beta^-_m}{\ell_1 \ell_2}
\left[\Psi_1(B_m^-,B_m^+)(1+\cos \sigma \cos \delta)
-\Psi_1 (B_m^+, B_m^-)\sin \sigma \sin \delta\right]
\right]{\dot x}-\Theta (\beta_m^+{\dot \sigma} + 
\beta_m^-{\dot \delta}),
\\
\\
\Theta \d2dot^{..} =
-2g\left[m\Psi_2 (-\ell^-, \ell^+)
-\dfrac{1}{\ell_1\ell_2}\Psi_1 (B_m^+,-B_m^-)\sin \delta\right]
-2k\Psi_1(\ell^-, \ell^+)x+\\
+ \Psi_1 (\ell^-,\ell^+)\left[
\Psi_2 (B_m^+, B_m^-)({\dot \sigma}^2+{\dot \delta}^2)
+2 \Psi_2 (B_m^-, B_m^+){\dot \sigma \dot \delta}
\right]+\\
+\Theta \left(\dfrac{f_1}{m_1\ell_1}-\dfrac{f_2}{m_2\ell_2}\right)-
\left[2\beta \Psi_1 (\ell^-,\ell^+)-m\Psi_1 
\left( \dfrac{\beta_2}{m_2 \ell_2}-\dfrac{\beta_1}{m_1 \ell_1}, 
\dfrac{\beta_1}{m_1 \ell_1}+\dfrac{\beta_2}{m_2 \ell_2}
\right)
\right.-\\
\left.
-\dfrac{2\beta^-_m}{\ell_1 \ell_2}
\left[\Psi_1(B_m^+,B_m^-)(1+\cos \sigma \cos \delta)
-\Psi_1 (B_m^-, -B_m^+)\sin \sigma \sin \delta\right]
\right]{\dot x}-\Theta (\beta_m^-{\dot \sigma} +  \beta_m^+{\dot \delta})
\end{array}
\right.
\end{equation}
where $\Theta= m - \dfrac{1}{2}\left[(m_1+m_2)+\Psi_1 (m_1+m_2,m_2-m_1)
\right]$ and 
\begin{equation}
\label{lpm}
\begin{array}{ll}
\ell^{\pm}=\dfrac{1}{2\ell_1\ell_2}(\ell_1 \pm \ell_2), & 
\beta_m^\pm = \dfrac{1}{2} 
\left(\dfrac{\beta_1}{m_1}\pm \dfrac{\beta_2}{m_2}\right).
\end{array}
\end{equation}
We notice that the escapement (\ref{escapement}) is missing in the first equation governing the beam's motion. 
Furthermore, even in the general case of different masses and lenghts the motion of $\delta$ (third equation) is more disentangled from the rest of the system than $\delta$: this property will be clearer afterward but actually can be prefigured here, if we imagine to replace in (\ref{eqlagrsdfn}) the functions(\ref{psi12}) with the second order Taylor polynomial
\begin{equation}
\label{psi12appr}
\begin{array}{ll}
\Psi_1 (\sigma, \delta; C_1, C_2)\approx C_1 - \dfrac{1}{4}[C_1(\sigma^2+\delta^2)-C_2 \sigma \delta], &
\Psi_2 (\sigma, \delta; C_1, C_2)\approx \dfrac{1}{2}(C_1\sigma+C_2\delta).
\end{array}
\end{equation}
\end{description}

\subsection{Equilibrium and stability}

\noindent
Clearly ${\bf q}={\bf 0}$ (i.~e.~$x=0$, $\theta_1=\theta_2=0$) is an equilibrium point for system (\ref{eqlagr}) if and only if $f_1=f_2=0$ (see (\ref{escapement})) at that position. In that case, since ${\bf y}={\bf 0}$ if and only if ${\bf q}={\bf 0}$, i.~e.~$x=0$, $\sigma=0$, $\delta=0$ is an equilibrium configuration also for system (\ref{eqlagrsd}) or (\ref{eqlagrsdfn}).

\noindent
Whenever ${\bf F}^{(\bf y)}+{\bm \Phi}^{(\bf y)}={\bf 0}$ (no escapement, no friction), the equilibrium point ${\bf q}={\bf 0}$ (hence also ${\bf y}={\bf 0}$) is Lyapunov stable by virtue of the Dirichlet's criterion, since it is an isolated minimum for the 
potential energy $V=\dfrac{1}{2}Kx^2-\sum\limits_{j=1}^2m_jg\ell_j \cos \theta_j$.
On the other hand, in presence of (\ref{damping1}) or (\ref{damping2}), the equilibrium at the same position is asimptotically stable: actually, we can write (\ref{damping1}) as 
$D({\bf q}){\dot {\bf q}}$, with $D={\footnotesize \left(\begin{array}{ccc} - \beta & 
-\beta_1 \ell_1 \cos \theta_1 & -\beta_2 \ell_2 \cos \theta_2 \\
-\beta_1 \ell_1 \cos \theta_1 & - \beta_1 \ell_1^2 & 0 \\
-\beta_2 \ell_2 \cos \theta_2 & 0 & -\beta_2 \ell_2^2
\end{array}
\right)}.$ Since $D$ is a negative--definite matrix, the energy balance $\dfrac{d}{dt}\left({\dot {\bf q}}\cdot 
\nabla_{\dot {\bf q}}{\cal L}-{\cal L}\right)={\dot {\bf q}}\cdot D{\dot {\bf q}}<0$
makes the energy a Lyapunov function tending to zero. In a similar way one can proceed for the case (\ref{damping2}).

\noindent
Nevertheless, if also the escapement (\ref{escapement}) is operating, it must be said that this does not connote stability (not even equilibrium) of the system, whichever force ${\bf F}^{({\bf q})}$ we make use of. 

\section{Elimination of the escapement. Absence of damping, friction}

\noindent
In this paper we are mainly involved in exploring the possibility of in--phase or antiphase synchronization settlements in absence of the escapement (\ref{escapement}): 
from now on, we will investigate the case ${\bf F}^{({\bf q})}={\bf 0}$
(i.~e.~no external devices are added to the system). Hence the terms containing $f_1$, $f_2$, $f^{pm}$ in (\ref{eqlagrsd}) or (\ref{eqlagrsdfn}) vanish. It must be sais that such as simplification entails a considerable advantage from the mathematical point of view, especially if (\ref{escapement}) is a step or discontinuous function, as it occurs in some mentioned models.

\noindent
In our first investigation even the friction forces are temporarily disregarded: the expected fact that synchronization is unattainable will find confirmation.

\noindent
Let us now examine the case when also $A_\beta^\pm$, $B_\beta^\pm$ (see (\ref{pm})) are removed: the equations of motion (\ref{eqlagrsd}) are now simply 
$\dfrac{d}{dt}\left( 
\nabla_{\dot {\bf y}}{\widetilde {\cal L}}\right)=
\nabla_{\bf y}{\widetilde {\cal L}}$ and, owing to the stability of the configuration ${\bf y}={\bf 0}$, we are ligitimated to replace (\ref{lagr}) with the quadratic expansion $\dfrac{1}{2}\left({\dot {\bf q}}\cdot {\bar A}{\dot {\bf q}}-{\bf q}\cdot {\bar V}{\bf q}\right)$, where 
${\bar A}={\footnotesize \left(\begin{array}{ccc} m& m_1\ell_1 & m_2 \ell_2 \\ m_1 \ell_1 & m_1\ell_1^2 & 0 \\
m_2 \ell_2 & 0 & m_2 \ell_2^2\end{array}\right)}$ and ${\bar V}=diag\,(k, m_1\ell_1 g, m_2 \ell_2 g)$. 
In terms of ${\widetilde {\cal L}}({\bf y}, {\dot {\bf y}})$, the linearized equations approximating (\ref{eqlagrsd}) are
$$
A_1\bfy2dot^{..}+V_1{\bf y}={\bf 0}, \quad
\begin{array}{cc}
A_1=L^{-1}{\bar A}L^{-1}={\footnotesize \left(\begin{array}{ccc}
m & B_m^+ & B_m^-\\ B_m^+ & A_m^+ & A_m^- \\ B_m^- & A_m^- & A_m^+
\end{array}
\right)}, &
V_1=L^{-1}{\bar V}L^{-1}=
{\footnotesize 
\left(\begin{array}{ccc}
k & 0 & 0 \\ 0 &  B_m^+g/2 &  B_m^-g/2 \\
0 & B_m^-g/2 &  B_m^+g/2
\end{array}
\right)}
\end{array}
$$ 
or explicitly
\begin{equation}
\label{eqlinnoattr}
\left\{ 
\begin{array}{l}
m\x2dot^{..} +
B_m^+\s2dot^{..}+B_m^-\d2dot^{..}=-kx,
\\
\\
B_m^+\x2dot^{..}+ A_m^+\s2dot^{..} + A_m^- \d2dot^{..} = -\dfrac{1}{2}g(B_m^+ \sigma + B_m^-\delta),
\\
\\
B_m^-\x2dot^{..}+A_m^{-}\s2dot^{..} + A_m^+\d2dot^{..} = -\dfrac{1}{2}g(B_m^-\sigma +  B_m^+ \delta).
\end{array}
\right.
\end{equation}

\noindent
Clearly, the same set of equations can be obtained directly from (\ref{eqlagrsd}), by replacing (\ref{psi12}) with the approximations (\ref{psi12appr}) and by neglecting all second order terms.
The fundamental frequencies (for both systems in ${\bf q}$ and ${\bf y}$) are found by solving $det\, (\lambda {\bar A}-{\bar V})=0$, leading to
\begin{equation}
\label{fundfreq}
(1-2\mu)\lambda^3- {\bar \lambda}\left[Y +2\Lambda^2\left(1-\mu +\rho\right)\right] \lambda^2+ {\bar \lambda}^2\Lambda^2 \left( 1 +2\gamma\right)\lambda - {\bar \lambda}^3\Lambda^2 Y =0
\end{equation}
with
\begin{equation}
\label{costanti}
\mu=\dfrac{m_1+m_2}{2m}, \quad
{\bar \lambda} = \dfrac{2g}{\ell_1+\ell_2}, 
\quad Y = \dfrac{k(\ell_1+\ell_2)}{2mg}, 
\quad \Lambda = \dfrac{\ell_1+\ell_2}{2\sqrt{\ell_1\ell_2}}, \quad \rho = \dfrac{\ell_1 - \ell_2}{\ell_1+\ell_2} \dfrac{m_1-m_2}{2m}
\end{equation}

\noindent
If the two lenghts $\ell_1=\ell_2=\ell$ are the same (but not necessarily the masses $m_1$ and $m_2$ are the same) it results $\Lambda=1$, $\rho=0$ and (\ref{fundfreq}) reduces to
\begin{equation}
\label{omega}
(1-2\mu)\lambda^3-\omega^2[Y+2(1-\mu)]\lambda^2+\omega^4(1+2Y)\lambda-\omega^6\gamma=0, 
\quad
\omega=\sqrt{\dfrac{g}{\ell}}.
\end{equation}
The quantity $\omega^2$ is an evident solution of it: this corresponds to the simplification of the third equation in (\ref{eqlagrsdfn}) to $\d2dot^{..}= -g\delta/\ell$. The remaining two solutions are 
(say $1$ with $-$ and $2$ with $+$)
\begin{equation}                                                      
\label{lambda12}
\omega^2_{1,2}=\dfrac{1+Y}{2(1-2\mu)}\omega^2
\left[
1 \pm \left(1-\dfrac{4Y}{(1+Y)^2}(1-2\mu)\right)^{1/2}\right]
\end{equation}

\noindent
We notice that $\omega_1\not = \omega_2$, since $(1+Y)^2>4Y (1-2\mu)$, being $\mu>0$. Moreover, both $\omega_1$ and $\omega_2$ are different from $\omega$, since $\mu<1/2$ and from the relation 
\begin{equation}
\label{relation}
\dfrac{\ell^2}{2g}\dfrac{( \omega_1^2 -\omega^2)( \omega_2^2 -\omega^2)}
{\omega_1^2 - \omega_2^2}=
\mu \ell\left( {1-\dfrac{4Y}{(1+Y)^2}(1-2\mu)}\right)^{-1/2}=B>0
\end{equation}
we deduce $\omega_1 < \omega$, $\omega_2 >\omega$.  
Finally, we remark that 
\begin{equation}
\label{limiti}
\begin{array}{lll}
\lim\limits_{\mu\rightarrow 0^+} \omega_1=\omega, & 
\lim\limits_{\mu\rightarrow 0^+} \omega_2=\sqrt{Y} \omega =\sqrt{k/m}, & \textrm{if}\;Y\geq 1\\
\lim\limits_{\mu\rightarrow 0^+} \omega_1=\sqrt{Y} \omega=\sqrt{k/m}, & 
\lim\limits_{\mu\rightarrow 0^+} \omega_2=\omega, & \textrm{if}\;\;0<Y<1 \\
\lim\limits_{\mu\rightarrow (1/2)^-} \omega_1=\left(\dfrac{Y}{1+Y}\right)^{1/2}\omega, & 
\lim\limits_{\mu\rightarrow (1/2)^-} \omega_2=+\infty & \textrm{for any}\;Y>0
\end{array}
\end{equation}
($\mu\approx 0$ means that the masses of the pendula are negligible with respect to the frame's one, on the contrary when $\mu\approx 1/2$ the mass of the frame is negligible).

\noindent
The generalized eigenvectors $(\lambda A_1 - V_1){\bf v}={\bf 0}$ corresponding to $\lambda=\omega_j^2$, $j=1,2$ and ${\bar \lambda}$ are respectively $(1, -2(\ell-g/\omega_j^2)^{-1}, 0)^T$, $j=1,2$ and $(0,(1-m_1/m_2)(1+m_1/m_2)^{-1}, 1)^T$ 
%$$
%\begin{array}{ccc}
%\left( 
%\begin{array}{c} 1 \\ 
%-\dfrac{2}{\ell- g/\omega_1^2} \\ 0 
%\end{array}
%\right), &
%\left( 
%\begin{array}{c} 1 \\ 
%-\dfrac{2}{\ell - g/\omega_2^2} \\ 0 
%\end{array}
%\right), &
%\left( \begin{array}{c} 0 \\ 
%\dfrac{1-m_1/m_2}{1 +m_1/m_2} \\ 1
%\end{array}
%\right)
%\end{array}
%$$
so that the solution ${\bf y}(t)$ starting from ${\bf y}(0)=(x(0), \sigma (0), \delta(0))^T$, ${\dot {\bf y}}(0)=({\dot x}(0), {\dot \sigma}(0), {\dot \delta}(0))^T$ is 
%$\left( 
%\begin{array}{c}
%x(0) \\ \sigma(0) \\ \delta(0) \end{array}
%\right)$, 
%${\dot {\bf y}}(0)=\left( 
%\begin{array}{c} 
%{\dot x}(0) \\ {\dot \sigma}(0) \\ {\dot \delta}(0) \end{array}
%\right)$ is

{\footnotesize 
\begin{eqnarray}
\nonumber 
x(t)&=& B\left[ 
\left( 2\zeta_2\dfrac{x(0)}{\ell}+C_0\right)^2+
\dfrac{1}{\omega_1^2}
\left( 2\zeta_2\dfrac{{\dot x}(0)}{\ell}+{\dot C}_0\right)^2\right]^{1/2}
\cos (\omega_1 t-\phi_2)-\\
\nonumber
&-& B\left[ 
\left( 2\zeta_1 \dfrac{x(0)}{\ell}+C_0\right)^2+
\dfrac{1}{\omega_2^2}
\left( 2\zeta_1 \dfrac{{\dot x}(0)}{\ell}+{\dot C}_0\right)^2
\right]^{1/2}\cos (\omega_2 t-\phi_1), 
\\
\label{soluzell}
%%%%%%%%%%%%%%%%%%%%%%%%%%%%%%%%%%%%%%%%%%%%%%%%%%%%%%%%%%%%%%%
\sigma(t)&=&
\dfrac{2B}{\ell}\left[ 
\left( \dfrac{Y}{\mu \ell} x(0) - \zeta_1 C_0\right)^2+
\dfrac{1}{\omega_1^2}
\left( \dfrac{Y}{\mu \ell} {\dot x}(0)-\zeta_1{\dot C}_0\right)^2 \right]^{1/2}\cos (\omega_1 t-\alpha_1)-\\
\nonumber
&-&
\dfrac{2B}{\ell}\left[ 
\left(
\dfrac{Y}{\mu \ell} x(0) - \zeta_2 C_0\right)^2
+
\dfrac{1}{\omega_2^2}
\left( \dfrac{Y}{\mu \ell} {\dot x}(0)-\zeta_2{\dot C}_0\right)^2
\right]^{1/2} \cos (\omega_2 t-\alpha_2)-\\
\nonumber
&-&
\dfrac{m_1-m_2}{m_1+m_2}\left( \delta^2(0)+\dfrac{{\dot \delta}^2(0)}{\omega^2}\right)^{1/2} \cos (\omega t-\alpha), \\
\nonumber
\delta(t)&=& 
\left( \delta^2(0)+\dfrac{{\dot \delta}^2(0)}{\omega^2}\right)^{1/2} \cos (\omega t-\alpha)
\end{eqnarray} }
where $B$ is defined in (\ref{relation}]) and
{\footnotesize 
$$
\begin{array}{ll}
\zeta_j=\dfrac{\omega_j^2}{\omega_j^2-\omega},\;\; 
C_0 = \sigma (0) + \dfrac{m_1 - m_2}{m_1+m_2} \delta (0), \;\;
{\dot C}_0 = {\dot \sigma} (0) + \dfrac{m_1 - m_2}{m_1+m_2} {\dot \delta} (0), 
\;\; \tan \alpha = \dfrac{{\dot \delta}(0)}{\omega \delta (0)} & \\
\tan \phi_j = \dfrac{1}{\omega_j^2}
\left( 2\zeta_j\dfrac{{\dot x}(0)}{\ell}
+{\dot C}_0\right)/\left( 2\zeta_j\dfrac{x(0)}{\ell}+C_0\right), & j=1,2\\
\tan \alpha_j= \dfrac{1}{\omega_j^2}
\left( \dfrac{Y}{\mu \ell} {\dot x}(0)-\zeta_j {\dot C}_0\right)/
\left( \dfrac{Y}{\mu \ell} x(0) - \zeta_j C_0\right), 
& j=1,2
\end{array}
$$
}

\noindent
We wrote explicitly the solution in order to remark that
\begin{description}

\item[1.] the motion of $\delta$ is independent either of $x$, $\sigma$ and of $\mu$, even if $m_1\not =m_2$: as a consequence, none of the initial sets lead to synchronization (i.~e.~$\delta(t)\rightarrow 0$), except for the matching of initial data ($\delta(0)=0$, ${\dot \delta}(0)=0$): in that case the pendula are exactly in--phase synchronized at any time.

\item[2.] The (not null) initial data $\delta(0)$ and ${\dot \delta}(0)$ produce an effect on the motion of $x$ and $\sigma$ if and only if the masses $m_1$ and $m_2$ are different.
\item[] The antiphase synchronization ($\sigma(t)\rightarrow 0$) never occurs, except for the trivial case $\sigma (0)=0$, $x(0)=0$, ${\dot x}(0)=0$, ${\dot \sigma}(0)=0$, $m_1=m_2$
(whereas the date for $\delta$ are not null, otherwise the system is at rest): for such data $\sigma(t)\equiv 0$ (antiphase at each tme) and the frame is at rest.

\item[3.] The solution $x(t)$ is periodic if and only if $\dfrac{\sqrt{Y(1-2\mu)}}{1+Y}=\dfrac{q}{1+q^2}$, for some rational number $q\in (0,1)$; if $m_1=m_2$, then $\sigma(t)$ is also periodic, otherwise it must be added the condition $\sqrt{\dfrac{Y}{1+Y}\dfrac{1+q^2}{q^2}}\in {\Bbb Q}$ in order to have $\sigma (t)$ periodic.
\item[4.] In order to have some understanding of the amplitudes of oscillation brought along each datum, we represent (\ref{relation}) by means of the parametric functions $B_\mu(Y)$, $Y\in (0,+\infty)$ with respect to the parameter $\mu\in (0,1/2)$: one can easily see that 
$$
\begin{array}{l}
B_{\mu_1}(Y)<B_{\mu_2}(Y)\;\; \textrm{if}\;\;0<\mu_1<\mu_2<1/2, \qquad
\lim\limits_{Y\rightarrow +\infty} B_\mu(Y)=0\;\textrm{for any}\;\mu\in (0,1/2),\\
0<\mu\leq 1/4:\;\;\textrm{the global maximum is}\;\;B_\mu(0)=\mu \ell \;\;\textrm{(strictly decreasing),} \\
1/4\leq \mu <1/2:\;\;\textrm{the global maximum is}\;\;B_\mu(1-4\mu)=\sqrt{\dfrac{\mu}{2(1-2\mu)}}\dfrac{\ell}{2}, 
\end{array}
$$
Consequently, the functions ${\mit \Phi}_\mu(Y)=\dfrac{2 B_\mu(Y)Y}{\mu\ell^2}$, ${\mit \Psi}^{(1)}_\mu(Y)=-\dfrac{2}{\ell}\dfrac{B_\mu(Y)\omega_1^2}
{\omega_1^2 - \omega^2}$ and ${\mit \Psi}^{(2)}_\mu(Y)=\dfrac{2}{\ell}\dfrac{B_\mu(Y)\omega_2^2}
{\omega_2^2 - \omega^2}$ where $\lambda_j(Y, \mu)$,  $j=1,2$, is (\ref{lambda12}), verify
$$
\begin{array}{l}
{\mit \Phi}_\mu(0)=0, \quad \lim\limits_{Y\rightarrow +\infty} {\mit \Phi}_\mu(Y)=2/\ell, \quad 
\textrm{for any}\;\mu\in (0,1/2), \quad
{\mit \Phi}_{\mu_1}(Y)>{\mit \Phi}_{\mu_2}(Y)\;\; \textrm{if}\;\;0<\mu_1<\mu_2<1/2, \\
0<\mu\leq 1/4:\;\;\textrm{the global maximum is}\;\;{\mit \Psi}_\mu(1/(1-4Y))= \dfrac{1}{\sqrt{\mu(1-2\mu)}}\dfrac{1}{2\ell}, \\
1/4\leq \mu <1/2:\;\;\textrm{the supremum is}\;\;2/ \ell  \;\;\textrm{(strictly increasing)} ,\\
{\mit \Psi}^{(1)}_\mu (0)=0,\quad {\mit \Psi}^{(1)}_\mu(1)=1/2, \quad
\lim\limits_{Y\rightarrow +\infty} {\mit \Psi}^{(1)}_\mu(Y)=1, \quad 
{\mit \Psi}^{(1)}_\mu \;\;\textrm{is strictly increasing}\;\;
\textrm{for any}\;\mu\in (0,1/2), \\
{\mit \Psi}^{(1)}_{\mu_1}(Y)<{\mit \Psi}^{(1)}_{\mu_2}(Y)\;\;[\textrm{resp.~}>]\;\; 
\textrm{if}\;\;0<\mu_1<\mu_2<1/2\;\;\textrm{and}\;\;0<Y<1\;\;[\textrm{resp.~}Y>1], \\
{\mit \Psi}^{(2)}_\mu (0)=1,\quad {\mit \Psi}^{(2)}_\mu(1)=1/2, \quad
\lim\limits_{Y\rightarrow +\infty} {\mit \Psi}^{(2)}_\mu(Y)=0, \quad 
{\mit \Psi}^{(2)}_\mu \;\;\textrm{is strictly decreasing}\;\;
\textrm{for any}\;\mu\in (0,1/2), \\
{\mit \Psi}^{(2)}_{\mu_1}(Y)>{\mit \Psi}^{(2)}_{\mu_2}(Y)\;\;[\textrm{resp.~}<]\;\; 
\textrm{if}\;\;0<\mu_1<\mu_2<1/2\;\;\textrm{and}\;\;0<Y<1\;\;[\textrm{resp.~}Y>1]. \\
\end{array}
$$
\end{description}

\noindent
As for the last point, the quantities $Y$ and $\mu$ are considered to be independent of each other by assuming, as an instance, the lenght $\ell$ and mass of the frame $m$ as fixed and modifying $m_1$, $m_2$ and $k$.

\noindent
Finally, we comment the case of different lenghts of the pendula.
The circumstance is in whole analogous to the previous case, except for the lesser simplicity of solving (\ref{fundfreq}), in order to achieve expressions similar to 
(\ref{soluzell}). With respect to (\ref{costanti}), we have $\Lambda=1$ if and only if $\ell_1=\ell_2$, whereas $\rho=0$ whether $\ell_1=\ell_2$ or $m_1=m_2$.
Hence, $\Lambda>1$ or $\rho\not =0$ causes somehow a ``disturbance'' to the exact solutions we found for $\ell_1=\ell_2$. 
In regards of that, we only point out that, calling $P(\lambda)$ the third--degree polynomial of (\ref{fundfreq}), we have
$P({\bar \lambda})=(1-\Lambda^2)(1-2\mu -Y)-2\Lambda^2\rho$. Let us have, for instance, $m_1\leq m_2$: assuming $\ell_1<\ell_2$, $Y>1-2\mu$ [respectively $\ell_1>\ell_2$, $Y<1-2\mu$ ], then $P({\hat \lambda})=0$ for  ${\hat \lambda}<{\bar \lambda}$ [resp.~$>$] (see (\ref{costanti})), i.~e.~the fundamental frequency decreases [resp.~increases] by comparison with the case $\ell_1=\ell_2$.
Similar remarks can be done about the other two solutions (\ref{lambda12}), for which it is
\begin{eqnarray}
\label{disturb}
P(\omega^2_{1,2})&=&4\left\{ [1-Y+2\mu Y(3+Y)](1-\Lambda^2)+2\rho[1-Y^2+2Y \mu]\Lambda^2\pm \right.\\
\noindent
&\pm&
\left.
[(1+2\mu Y)(1-\Lambda^2)+2\rho (1+Y)\Lambda^2)]\left[(1+Y^2)-4Y(1-2\mu)\right]^{1/2}\right\}
\end{eqnarray}

\section{Elimination of the escapement. Presence of damping, friction}

\noindent
Let us start from system (\ref{eqlagrsdfn}): defining $y={\dot x}$, $u={\dot \sigma}$, $v={\dot \delta}$ and setting the system as ${\dot {\bf x}}={\cal F}({\bf x})$
with ${\bf x}=(x,\sigma, \delta, y, u,v)^T$, we consider the linear approximation ${\dot {\bf x}}=(J_{\bf x}{\cal F}\vert_{{\bf x}={\bf 0}}){\bf x}$, where $J_{\bf x}$ calculates the Jacobian matrix. Eliminating the terms containing the escapement $f_1$, $f_2$ and reverting to the explicit expressions of the constant quantities (\ref{pm}), one gets

\begin{equation}
\left\{
\label{fnlinnof}
\begin{array}{l}
{\dot x}=y, \qquad {\dot \sigma} = u, \qquad {\dot \delta}=v, \\
{\dot y}=\dfrac{1}{m(1-2\mu)}\left( -kx+\dfrac{1}{2}(m_1+m_2) g \sigma +\dfrac{1}{2}
(m_1-m_2) g \delta -\beta_0 y\right), \\
\\
{\dot u}=\dfrac{1}{m(1-2\mu)\ell_1 \ell_2}\left\{ (\ell_1 + \ell_2)\left(
kx-m\dfrac{g}{2} \sigma\right)+
\dfrac{g}{2}\left[m(\ell_1-\ell_2)-2(m_1\ell_1-m_2\ell_2)\right]
\delta +\right.\\
\left. + 
\left[ (\ell_1+\ell_2)\beta
-m\left(\dfrac{\beta_1}{m_1}\ell_2 +\dfrac{\beta_2}{m_2}\ell_1 \right)
-\left(\dfrac{\beta_1}{m_1}- \dfrac{\beta_2}{m_2} \right)
(m_1\ell_1 - m_2\ell_2)
\right] y\right\}- \\
-\dfrac{1}{2}\left(\dfrac{\beta_1}{m_1}+\dfrac{\beta_2}{m_2}\right) u 
-\dfrac{1}{2}\left(\dfrac{\beta_1}{m_1}-\dfrac{\beta_2}{m_2}\right) v,
\\
\\
{\dot v}= \dfrac{1}{m(1-2\mu)\ell_1 \ell_2}\left\{ (\ell_1-\ell_2)
\left(-kx+m\dfrac{g}{2} \sigma\right)-
\dfrac{g}{2}\left[m(\ell_1+\ell_2)-2(m_1\ell_1+m_2\ell_2)\right]\delta +\right.\\
\left.
+\left[-(\ell_1-\ell_2)\beta
-m\left(\dfrac{\beta_1}{m_1}\ell_2- \dfrac{\beta_2}{m_2}\ell_1 \right)
+\left(\dfrac{\beta_1}{m_1}- \dfrac{\beta_2}{m_2} \right)
(m_1\ell_1 + m_2\ell_2)
\right]y\right\}-\\
-\dfrac{1}{2}\left(\dfrac{\beta_1}{m_1}-\dfrac{\beta_2}{m_2}\right) u 
-\dfrac{1}{2}\left(\dfrac{\beta_1}{m_1}+\dfrac{\beta_2}{m_2}\right) v
\end{array}
\right.
\end{equation}

\begin{rem}
In the presence of escapement (\ref{escapement}), the terms
$\sum\limits_{j=1}^2\dfrac{1}{m_j\ell_j}\left( f_{j,\sigma}\sigma+f_{j,\delta}\delta+f_{j,u}u+f_{j,v}v\right)$ and $\dfrac{1}{m_1\ell_1}\left( f_{1,\sigma}\sigma+f_{1,\delta}\delta+f_{1,u}u+f_{1,v}v\right)-
\dfrac{1}{m_2\ell_2}\left( f_{2,\sigma}\sigma+f_{2,\delta}\delta+f_{2,u}u+f_{2,v}v\right)$
where $f_{i,\zeta}=\dfrac{\partial f_i}{\partial \zeta}$, $i=1,2$, $\zeta=\sigma, \delta, u,v$, are calculated at the equilibrium ${\bf x}={\bf 0}$, must be added to the fifth
and sixth equations, respectively. 
\end{rem}

\noindent
This time, the motion of $\delta$ (last equation in (\ref{fnlinnof})) is not independent of the other variables if simply $\ell_1=\ell_2$: actually, this equation is completely disentangled from the rest when also $m_1=m_2$, $\beta_1=\beta_2$. 

\subsection{Localization of the eigenvalues}

\noindent
The characteristic polynomial associated with the linear system (\ref{fnlinnof}) is 
\begin{equation}
\label{polcarattr}
\begin{array}{l}
(1-2\mu)\lambda^6 + 
\left\{ \dfrac{\beta_0}{m} + (1-2\mu)
\left( \dfrac{\beta_1}{m_1}+\dfrac{\beta_2}{m_2}\right)\right\} \lambda^5+\\
+\left\{\dfrac{\beta_0}{m} \left( \dfrac{\beta_1}{m_1}+\dfrac{\beta_2}{m_2}\right)
+(1-2\mu) \dfrac{\beta_1}{m_1}\dfrac{\beta_2}{m_2} +
\dfrac{g}{\ell_2} \dfrac{m-m_1}{m} +\dfrac{g}{\ell_1} \dfrac{m-m_2}{m} +\dfrac{k}{m} \right\} \lambda^4+\\
+\left\{ \dfrac{\beta_1}{m} \left( \dfrac{g}{\ell_2}\dfrac{m-m_1}{m_1}+\dfrac{g}{\ell_1}\right) + 
\dfrac{\beta_2}{m} \left( \dfrac{g}{\ell_1}\dfrac{m-m_2}{m_2}+\dfrac{g}{\ell_2}\right)
 + \dfrac{k}{m} \left(  \dfrac{\beta_1}{m_1}+\dfrac{\beta_2}{m_2}\right)\right\} \lambda^3+\\
+\left\{ 
\dfrac{g^2}{\ell_1\ell_2} + 
\dfrac{g}{\ell_1}
\left(
\dfrac{k}{m}+\dfrac{\beta_1}{m_1}\dfrac{\beta_0 + \beta_2}{m}
\right)
+\dfrac{g}{\ell_2}
\left(
\dfrac{k}{m} +\dfrac{\beta_2}{m_2} \dfrac{\beta_0 + \beta_1}{m}
\right)
\right\} 
\lambda^2+\\
+\left\{
\dfrac{g^2}{\ell_1\ell_2} \dfrac{\beta}{m}+\dfrac{k}{m}\left( 
\dfrac{g}{\ell_2}\dfrac{\beta_1}{m_1}
+\dfrac{g}{\ell_1}\dfrac{\beta_2}{m_2} \right)\right\} \lambda+
\dfrac{g^2}{\ell_1\ell_2}\dfrac{k}{m}=0.
\end{array}
\end{equation}
The terms with odd exponent of $\lambda$ are due only to the friction contributions (\ref{damping1}): actually, when $\beta_j=0$ for each $j=0,1,2$, the characteristic problem  (\ref{polcarattr}) is equivalent to the one formulated in (\ref{fundfreq}). 

\noindent
Besides the presence of a Liapunov function which guarantees the asymptotical stability (see Par.~2.1), it is worthwhile to assert also the following  

\begin{propr}
The real part of each root of the polynomial (\ref{polcarattr}) is negative.
\end{propr}  

\noindent
{\bf Proof.} It is sufficient to make use of the 
Routh--Hurwitz criterion (see, for istance, \cite{gan}): writing (\ref{polcarattr}) as $\sum\limits_{n=1}^6 a_n \lambda^n=0$, $a_6=1-2\mu$, $\dots$, $a_0=\dfrac{g^2}{\ell_1\ell_2}\dfrac{k}{m}$ it can be checked, even though calculations last long, that the chain of seven numbers required for the mentioned criterion
$$
\begin{array}{l}
a_6, \;\;\; a_5, \;\;\;\dfrac{1}{a_5}b_1,\; \;\; a_3-a_5\dfrac{b_2}{b_1}, \\
\dfrac{1}{a_5} \left(b_2 - b_1 \dfrac{a_1b_1-a_0}{a_3b_1-a_2b_2}\right), 
\;\;\; a_1-\dfrac{a_0}{b_1}-a_0a_5\left( a_3 - \dfrac{b_2}{b_1}a_5\right)
\left(b_2-b_1\dfrac{a_1b_1 - a_0}{a_3b_1-a_2b_2}\right)^{-1}, 
\;\;\; a_0, 
\end{array}
$$
where $b_1=a_4a_5-a_3a_6$, $b_2=a_2a_5-a_1a_6$,  consists of all positive numbers. 
Since the sequence has no sign change, all the roots of the sixth degree polynomial (\ref{polcarattr}) have negative real parts. $\quad\square$

\noindent
The criterion places the eigenvalues in the left half plane of the complex plane. 
In order to discriminate the occurrences of real roots, or conjugate pairs of complex roots of (\ref{polcarattr}) one way could be calculate the discriminant which gives additional information on the nature of the roots, real or complex, although calculations for the sixth degree polynomial (\ref{polcarattr}) are quite complex.

\noindent
Our definitive aim is to infer some information about the qualitative behaviour of system 
(\ref{fnlinnof}), by means of locationing as much as possible the portion on the complex plane where the solutions of (\ref{polcarattr}) lie. 

\begin{rem}
It must be said that the Gershgorin circle theorem used to bound the spectrum 
is not, in our case, especially powerful: it can be easily seen that the Gershgorin discs where the eigenvalues are confined 
do not keep them away from the origin of the complex plane: this will be a crucial point in our analysis.
\end{rem}

\subsubsection{The case of identical pendula}

\noindent
Equation (\ref{polcarattr}) will be now discussed in the simpler case of identical pendula: we will assume from now on
\begin{equation}
\label{idpend}
\begin{array}{lll}
m_1=m_2=m_p, & \ell_1 =\ell_2=\ell, & \beta_1=\beta_2=\beta_p
\end{array}
\end{equation}
(the subscript $p$ is necessary in order not to confuse with the quantities defined in (\ref{lagr}) and (\ref{damping1})).
Nevertheless, we presume that most of the shown results are still valid in the general case of different pendula.

\noindent
We avoid to write again system (\ref{fnlinnof}) in case of assumption (\ref{idpend}), since the simplifications are evident. 
As we touched upon, assumption (\ref{idpend}) makes the motion of $\delta$ independent of the rest of the system: in fact, the most evident advantage of (\ref{idpend}) is the reduction of the equation for $\delta$ simply to 
$\d2dot^{..}+\dfrac{\beta_p}{m_p} {\dot \delta}+\dfrac{g}{\ell}\delta=0$ with negative eigenvalues  
\begin{equation}
\label{eta}
\begin{array}{lllll}
-\dfrac{1}{2} \left(\eta \pm \sqrt{\eta^2-4}\right)\omega &\textrm{if}\;\;\eta^2\geq 4, &
-\dfrac{1}{2} \left(\eta \pm i\sqrt{4-\eta^2}\right)\omega &\textrm{if}\;\;\eta^2< 4, &
\eta=\dfrac{1}{\omega}\dfrac{\beta_p}{m_p}=\dfrac{\beta_p}{m_p}\sqrt{\dfrac{\ell}{g}} 
\end{array}
\end{equation}
giving the solution 
\begin{equation}
\label{soluzdelta}
\delta(t)=
\left\{
\begin{array}{ll}
e^{-\dfrac{1}{2} \eta \omega t}\left( \delta(0)
\cosh \left(\dfrac{1}{2} \sqrt{\eta^2-4} \omega t \right)+ 
\dfrac{2{\dot \delta}(0)}{\sqrt{\eta^2-4}\omega}\sinh 
\left(\dfrac{1}{2} \sqrt{\eta^2-4} \omega t \right)\right)& \textrm{if}\;\;\eta^2>4\\
e^{-\dfrac{1}{2} \eta \omega t}\left( \delta(0)\cos 
\left(\dfrac{1}{2} \sqrt{4-\eta^2} \omega t \right)+\dfrac{2{\dot \delta}(0)+\eta \omega
\delta(0)}{\sqrt{4-\eta^2} \omega}\sin \left(\dfrac{1}{2} \sqrt{4-\eta^2} \omega t \right)\right)
& \textrm{if}\;\;\eta^2<4\\
e^{-\dfrac{1}{2} \eta \omega t}\left(\delta(0)+\left(
{\dot \delta}(0)+\dfrac{1}{2} \eta \omega \delta (0)\right)t\right)& \textrm{if}\;\;\eta^2=4
\end{array}
\right.
\end{equation}
The characteristic equation (\ref{polcarattr}) can be now written as 
\begin{equation}
\label{polcarip}
\begin{array}{l}
\left( \lambda^2+\dfrac{\beta_p}{m_p}\lambda+\dfrac{g}{\ell}\right)\left[
(1-2\mu)\lambda^4+
\left(\dfrac{\beta_0}{m} +(1-2\mu)\dfrac{\beta_p}{m_p}\right)\lambda^3+\left(\dfrac{k}{m} +
\dfrac{\beta_0}{m}\dfrac{\beta_p}{m_p}+\dfrac{g}{\ell}\right)\lambda^2+\right.\\
\left.+
\left(\dfrac{k}{m} \dfrac{\beta_p}{m_p}
+\dfrac{\beta_0+2\beta_p}{m}\dfrac{g}{\ell} \right)\lambda+
\dfrac{k}{m} \dfrac{g}{\ell}\right]=0
\end{array}
\end{equation}
where the factorization is related to the uncoupling of $\delta$ from the system. We will focus our attention on the factor between square brackets, which gives the eigenvalues 
related to the motion of $x$ and $\sigma$.

\noindent
If, in addition to $\eta$ (see (\ref{eta})), $\omega$ (see (\ref{omega})), $Y=\dfrac{k}{m}\dfrac{\ell}{g}$ (see (\ref{costanti})), we define 
\begin{equation}
\label{x}
X=\dfrac{\beta_0}{m}\sqrt{\dfrac{\ell}{g}},
\end{equation}
then the fourth degree polynomial in square brackets, eq.~(\ref{polcarip}), can be written as
\begin{equation}
\label{polcarip4}
(1-2\mu)\lambda^4+[X+(1-2\mu)\eta]\omega \lambda^3+(\eta X+Y+1)\omega^2\lambda^2+
(\eta Y +X + 2\mu \eta ) \omega ^3 \lambda+Y \omega^4=0
\end{equation}
We notice that $\eta$, $X$ and $Y$ are adimensional quantities (whereas the units of $\omega$ are $time^{-1}$).

\noindent
At this point, we employ the Enestr\"om--Kakeya Theorem (\cite{ene}, \cite{kak}), in the following version:

\begin{propr} {\rm (E--K Theorem)} Let $p_n(\lambda)=a_0 + a_1 \lambda + \dots + a_{n-1}\lambda^{n-1}+a_n\lambda^n$ a polynomial with 
$a_j>0$ for any $j=0,\dots, n$. Then, all the zeros of $p_n$ are contained in the annulus
of the complex $z$--plane
\begin{equation}
\label{annulus}
\rho_m\leq |z| \leq \rho_M, \quad 
\rho_m=\min\left\{ \dfrac{a_0}{a_1}, \dfrac{a_1}{a_2}, \dots, \dfrac{a_{n-2}}{a_{n-1}}, \dfrac{a_{n-1}}{a_n}\right\}, \;\;
\rho_M=\max\left\{ \dfrac{a_0}{a_1}, \dfrac{a_1}{a_2}, \dots, \dfrac{a_{n-2}}{a_{n-1}}, \dfrac{a_{n-1}}{a_n}\right\}
\end{equation}
\end{propr}

\noindent
By writing (\ref{polcarip4}) as $\sum\limits_{n=1}^4 a_n\lambda^n=0$ (even though the coefficients are different from the one used in the proof of Property $4.1$), the quantities needed for (\ref{annulus}) are
\begin{equation}
\label{rapporti}
\begin{array}{llll}
\dfrac{a_0}{a_1}=\dfrac{Y}{X+\eta Y+2\mu \eta}\omega, &
\dfrac{a_1}{a_2}=\dfrac{X+\eta Y+2\mu \eta}{\eta X+Y+1}\omega, &
\dfrac{a_2}{a_3}=\dfrac{\eta X+Y+1}{X+(1-2\mu)\eta}\omega, &
\dfrac{a_3}{a_4}=\dfrac{X+(1-2\mu)\eta}{1-2\mu}\omega.
\end{array}
\end{equation}
It is immediate to check that $\dfrac{a_0}{a_1}\leq \dfrac{a_2}{a_3}$ and 
$\dfrac{a_1}{a_2}\leq \dfrac{a_3}{a_4}$
for any data $\eta$, $\omega$, $X$ and $Y$, hence
\begin{equation}
\label{couple}
\rho_m=\min\left\{ \dfrac{a_0}{a_1},\dfrac{a_1}{a_2}\right\}, \quad 
\rho_M=\max\left\{\dfrac{a_2}{a_3},\dfrac{a_3}{a_4}\right\}
\end{equation}
and the positive quadrant ${\cal Q}=\{ (X,Y)\,|\,X>0,Y>0\}$ is splitted in the four regions 
\begin{equation}
\label{regioniz}
\begin{array}{ll}
{\cal Z}_1 = \left\{ (X,Y)\in {\cal Q}\,|\, \rho_m=\dfrac{a_0}{a_1}, 
\rho_M=\dfrac{a_2}{a_3}\right\} & 
{\cal Z}_2 = \left\{ (X,Y)\in {\cal Q}\,|\, \rho_m=\dfrac{a_0}{a_1}, 
\rho_M=\dfrac{a_3}{a_4}\right\} \\
{\cal Z}_3 = \left\{ (X,Y)\in {\cal Q}\,|\, \rho_m=\dfrac{a_1}{a_2}, 
\rho_M=\dfrac{a_2}{a_3}\right\} & 
{\cal Z}_4 = \left\{ (X,Y)\in {\cal Q}\,|\, \rho_m=\dfrac{a_1}{a_2}, 
\rho_M=\dfrac{a_3}{a_4}\right\}
\end{array}
\end{equation}

\noindent
Sketching a physical reading, we see that a state $(X,Y)\in {\cal Q}$ on the right part of the quadrant ($X\gg Y$) exhibits a predominance of the damping force on the elastic force, both acting on the beam. On the contrary, in the left upper part of the quadrant ($Y\gg X$) the effects are reversed.

\subsubsection{Locating the synchronization regions}

\noindent
If one opts for the point of view of fixing $\omega$ (lenght of pendula) and $\eta$ (damping of pendula) and let $X$ (friction of the beam) and $Y$ (elastic constant of the spring) vary, one can depict the four regions ${\cal Z}_i$, $i=1,2,3,4$ on the quadrant ${\cal Q}$.
In finding them, we see that the comparison of (\ref{rapporti}) leads to the following 
quadratic conditions in the variables $X$ and $Y$
\begin{equation}
\label{coniche}
\begin{array}{lll}
a_0/a_1<a_1/a_2 & \textrm{if and only if} 
& X^2+(\eta^2-1)Y^2+\eta X Y + 4 \mu \eta X + (4\mu \eta^2-1)Y + 4\mu^2\eta^2>0, \\
a_0/a_1<a_3/a_4 & \textrm{if and only if} 
& X^2+\eta XY + X + (1-2\mu)(\eta^2-1)Y +2 \mu (1-2\mu)\eta^2>0, \\
a_1/a_2<a_2/a_3 & \textrm{if and only if} 
&(\eta^2-1)X^2+Y^2+\eta XY +\eta X +[2-(1-2\mu)\eta^2]Y+\\
& & +1-2\mu (1-2\mu)\eta^2>0, \\
a_2/a_3<a_3/a_4 & \textrm{if and only if} 
&X^2+(1-2\mu)\eta X - (1-2\mu)Y+(1-2\mu)[(1-2\mu)\eta^2-1]>0
\end{array} 
\end{equation}
involving the construction of arcs of conics. 
It is easy to check that for $\eta^2>4/3$ the first three conditions define three regions 
on the $(X,Y)$ positive quadrant delimited by hyperbolae, for $\eta^2<4/3$ the first and the third conics are real ellipses ($\eta^2=4/3$: two intersecting lines). The fourth condition refers to a parabola attaining its vertex for some $X<0$.
The case $\eta\leq 1$, which will be examined deeper, is plotted in Figure 2, where the curves are numbered in the same order as in (\ref{coniche}).

\begin{figure}[htbp]
\includegraphics[scale=.30]{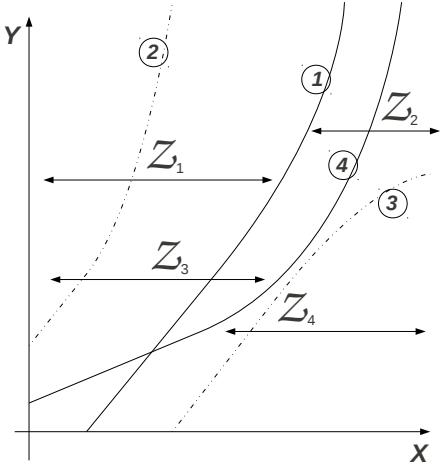}
\caption{The numbered curves are the conics appearing in (\ref{coniche}), in the same order. The minimum radius $\rho_m$ and the maximum radius $\rho_M$ delimiting the annulus which contains the spectrum are deduced from the inclusion of the status $({\bf X},{\bf Y})$ to one of the regions ${\cal Z}_i$, $i=1,2,3,4$, defined in (\ref{regioniz}). The dotted curves $2$ and $3$ refer only to the intermediate values in (\ref{rapporti}) between the minimum and the maximum.}
\end{figure}

\noindent
We make use now of the localization carried out by the E--K Theorem in order to 
compare the eigenvalues governing $\delta$ (see (\ref{eta})) and those governing $\sigma$ (solutions of (\ref{polcarip})). We will hereafter focus on the case $\eta\leq 1$, which is physically more consistent, asserting that the complementary case can be conceptually treated in the same way.

\noindent
Having in mind (\ref{eta}), we are in the case of conjugate complex eigenvalues for $\delta$ with $-\dfrac{1}{2}\eta \omega$ as real part and $\omega$ as modulus.
The main question is how the eigenvalues (\ref{eta}) are located with respect to 
the annulus (\ref{annulus}) delimited by the radii (\ref{couple}). 
We prove the following

\begin{prop}
For $\eta\leq 1$, the two roots (\ref{eta}) cannot lie in the half--plane $Re\;z<-\rho_M$.
\end{prop}

\noindent
{\bf Proof.} The real part of the two roots (\ref{eta}) is $-\dfrac{1}{2}\eta \omega$: 
they belong to the half--plane $Re\;z<\rho_M$ if and only if (see also (\ref{rapporti}))
$\eta > 2\dfrac{\eta X+Y+1}{X+(1-2\mu)\eta}$  in ${\cal Z}_1 \cup {\cal Z}_3$, 
$\eta > 2\dfrac{X+(1-2\mu)\eta}{1-2\mu}$  in ${\cal Z}_2 \cup {\cal Z}_4$.
However, the two conditions define empty regions in ${\cal Q}$, since they are equivalent to $Y+\dfrac{1}{2} \eta X < \dfrac{1}{2} (1-2\mu)\eta^2-1<0$ and $X<-\dfrac{1}{2} (1-2\mu)\eta<0$, respectively (we recall that $\mu <1/2$, see (\ref{costanti})).
$\quad \square$

\begin{rem} The eigenvalues (\ref{eta}) belong to the semicircumference $|z|=\omega$, $Re\,z <0$: the more specific question whether they lie in the semicircle $|z|\leq \rho_M$, $Re\;z>0$ can be easily solved, by
comparing $\omega$ with (\ref{couple}), second couple. It results that (\ref{eta}) are within the semicircle if and only if $X>(1-2\mu)(1-\eta)$ when $\rho_M= a_2/a_3$ (see (\ref{rapporti}) and if only if $Y>(1-\eta)X+(1-2\mu)\eta -1$ when $\rho_M=a_3/a_4$. 
Graphically, each of the two regions ${\cal Z}_1\cup {\cal Z}_3$ and 
${\cal Z}_2\cup {\cal Z}_4$
(see Figure 4) is splitted by a straight line and the required condition is true only on one side.  
\end{rem}

\begin{figure}[htbp]
\includegraphics[scale=.39]{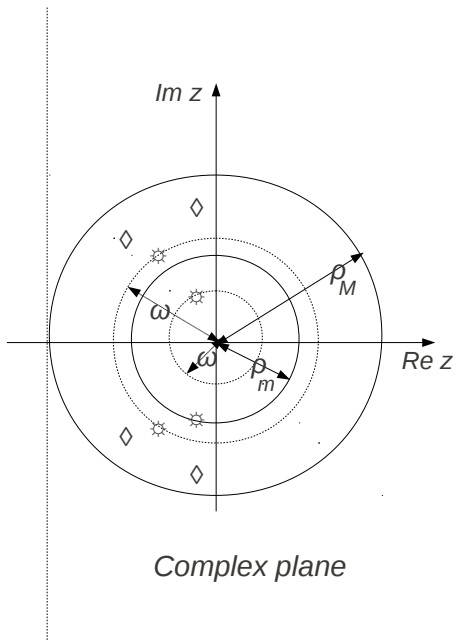}
\caption{Thepoints marked with $\diamond$ are the four solutions of (\ref{polcarip}), here considered as two couples of conjugate complex roots. The radius $\omega$ cannot exceed $\rho_M$: two possible values of $\omega$ are drawn, the smaller one $\omega<\rho_m$ referring to a state which facilitates the antiphase synchronization. }
\end{figure}

\noindent
Our first conclusion is that the system cannot establish a status where the difference $\delta(t)$ decays to zero more rapidly than the sum $\sigma(t)$: thus, the in-phase synchronization onset is inhibited.  
The result is consistent with the experimental detection, starting from Huygens, on the grounds that the antiphase synchronization is indeed prevailing on the inphase one in this sort of phenomenon. 

\noindent
We finally discuss the possibility for the system of establishing 
a status of  antyphase synchronization, still keeping $\eta\leq 1$.
Making use once again of the localization (\ref{couple}), we compare the real part of  (\ref{eta}) with $\rho_m$: whenever $\eta \omega <2\rho_m$, the decay of $\sigma$ is expected to be faster than the one of $\delta$, so that antyphase synchronization is facilitated. Thus, we are going to check whether (see \ref{rapporti}))
\begin{equation}
\label{etarm}
\begin{array}{llllll}
(A)&\eta < 2\dfrac{Y}{X+\eta Y + 2\mu \eta}&  in {\cal Z}_1 \cup {\cal Z}_2, & 
(B)&\eta < 2\dfrac{X+\eta Y+2\mu \eta}{\eta X+Y +1}&  in {\cal Z}_3 \cup {\cal Z}_4.
\end{array}
\end{equation}
The two conditions in (\ref{etarm}) define two half--planes above the straight lines
$(2-\eta^2)X+\eta Y - \eta (1-4\mu)=0$ and $\eta X + (\eta^2-2)Y+2\mu \eta^2=0$, respectively.

\noindent
As for condition $(A)$ of (\ref{etarm}), it can be easily seen that
\begin{description}
\item[] if $\mu \leq \dfrac{1}{2} \dfrac{2-\eta^2}{4-\eta^2}$ then 
$(A)$ is valid in 
${\cal Z}_1 \cup {\cal Z}_2$ except for a triangular lower region 
cut off by the straight line,  (see Figure 4)
\item[] if $\dfrac{1}{2} \dfrac{2-\eta^2}{4-\eta^2}<\mu<\dfrac{1}{2}$ then 
$(A)$ is valid in all the set.
\end{description}
By recalling $\mu=m_p/m$, we see that, $\eta$ being equal, the smaller is the mass of the pendula with respect to the mass of the system, the larger is the region which excludes the possibility of antyphase synchronization, i.~e.~the removed zone.

\noindent
Condition $(B)$ of (\ref{etarm}) eliminates all the lower part from ${\cal Z}_3 \cup {\cal Z}_4$ and selects the region bounded from below by the straight line 
$\eta X + (\eta^2-2)Y+2\mu \eta^2=0$ and 
from above by the parabola related to the fourth condition in (\ref{coniche}). Moreover
\begin{description}
\item[] if $\mu \geq \dfrac{1}{2} \dfrac{2-\eta^2}{4-\eta^2}$ then the selected region 
is bounded at the left hand side by a segment on the $Y$--axis, 
\item[] if $\mu <\dfrac{1}{2} \dfrac{2-\eta^2}{4-\eta^2}$ then the selected region  
is disjointed from the $Y$--axis. 
\end{description}
The different cases are summarized in Figure 4.

\begin{figure}[htbp]
\includegraphics[scale=.39]{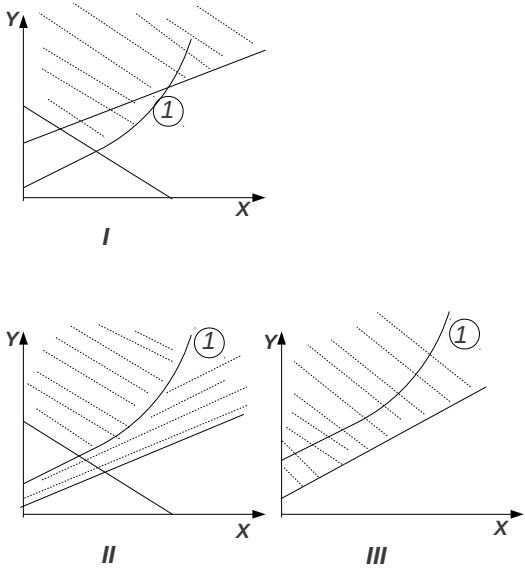}
\caption{The selection on the quadrant ${\cal Q}$ is refined by comparing $\omega$ with $\rho_m$. Picture $I$ refers to the case $\mu\leq \dfrac{1}{2} \dfrac{2-\eta^2}{4-\eta^2}$, picture $II$ to $\dfrac{1}{2} \dfrac{2-\eta^2}{4-\eta^2}< \mu<  \dfrac{1}{4}$ and picture $II$ to $\dfrac{1}{4} \leq \mu<\dfrac{1}{2}$. }
\end{figure}

\noindent
Let us call ${\cal A}\subset {\cal Q}$ the subset where conditions $(A)$ and $(B)$ of (\ref{etarm}) are both fulfilled.
Whenever a state  $(X,Y)\in {\cal A}$ is such that the four roots of (\ref{polcarip4}) are all real, then they exceed in modulus $\dfrac{1}{2} \eta\omega$, therefore the 
decay of $\delta(t)$ is lengthier than the decay of $\sigma$ and the system shows a tendency to an antiphase arrangement. 

\noindent
However, if some or all of the roots of (\ref{polcarip4}) are not real, the condition 
$(X,Y)\in {\cal A}$ does not guarantee by itself that the real part of such solutions are greater than $\dfrac{1}{2} \eta \omega$ in modulus. In order to overcome this problem, let
us explain a method, without delving into the detail: whenever the solutions of (\ref{polcarip4}) are, as an instance, all complex, one can start from the conditions
\begin{equation}
\label{condre}
(Re\,\lambda)^2+\dfrac{1}{2} \dfrac{a_3}{a_4} Re\,\lambda +\dfrac{1}{4} \left(
\dfrac{a_2}{a_4} -2\rho_m^2 \right)\geq 0
\end{equation}
where $a_2$, $a_3$ and $a_4$ are the same as in (\ref{rapporti}) and $Re\,\lambda$ is the real part of any root of (\ref{polcarip4}). 
The estimation (\ref{condre}) can be proved by setting the complex roots as
$\xi_r\pm i \kappa_r$, $r=1,2$, making use of the known relations between roots and coefficients: 
$\xi_1^2+\kappa_1^2+\xi_2^2+\kappa_2^2+ 4\xi_1 \xi_2 = a_2/a_4$, $2(\xi_1+\xi_2)=-a_3/a_4$ 
and finally recalling that $\xi_r^2+\kappa_r^2\geq \rho_m^2$, $r=1,2$. 
In a mixed occurrence of two real roots and two complex conjugate roots one argues in a similar way.

\noindent
Condition (\ref{condre}) can be used in order to separate the real parts away from zero: actually, if we require $\dfrac{a_2}{a_4}\leq 2\rho_m^2$, then $Re\,\lambda\leq B$, being 
$B$ the negative solution of (\ref{condre}) where $=$ replaces $\geq$.
Finally, by demanding $-\dfrac{1}{2}\eta \omega\geq B$, we achieve that the real parts of the solutions of (\ref{polcarip4}) exceed in modulus $\dfrac{1}{2} \eta \omega$. 
The conditions $\dfrac{a_2}{a_4}\leq 2\rho_m^2$ and $-\dfrac{1}{2}\eta \omega\geq B$, once they have been expressed in terms of $X$ and $Y$ via (\ref{rapporti}) and (\ref{couple}), 
will determine the regions of ${\cal Q}$ where antiphase synchronization is facilitated.

\section{Conclusions}

\noindent
In the frame of the study of coupled oscillations of two pendula, we intended to
pursue a double objective:
\begin{itemize}
\item[$1.$] to formulate the model in the experimental context as general as possible, 
\item[$2.$] to develop the corresponding mathematical problem in a simplified case, 
drawing the attention to the fact that some classical results about the localization of the spectrum of a matrix can allow us to predict the qualitative behaviour of the system.
\end{itemize}

\noindent
We selected the parameters $X$ (see (\ref{x})) and $Y$ (see (\ref{costanti})) in order to plot on the quadrant ${\cal Q}$ a certain number of regions in each of which the system 
develops in a different way. The parameters $\mu$ (definrd in (\ref{costanti})) and $\eta$ (see (\ref{eta}) are considered as constant, but different choices can be made in order to  
represent the states (as, for instance, fixing the friction of the borad $\beta_0$ but varying the damping of the pendula $\beta_p$).

\noindent
The sections of the complex plane where the spectrum is confined and the regions on ${\cal Q}$ are not computed via a numerical simulation but they are predicted by the analysis of the spectrum of the linearized system (\ref{fnlinnof}).

\noindent
Within specific ranges of the parameters, the tendendy of the system to evolve towards the antiphase synchronization, rather than the inphase one, is predicted by the present analysis. This conclusion is in step with the real development of the phenomenon, starting from the Huygens' observation in the $XVII$ century of the $180^°$ out of phase swings (see also \cite{fra})).

\noindent
Generally speaking, our purpose is to highlight that the method, beyond the specific 
circumstance which has been exerted to, can be extended to more general situations
where the role of certain parameters are exchanged or some restrictions are relaxed. 
At the same time, the analysis is appropriated, in our mind, to be combined with a simple numerical approach.

\noindent
Both the mentioned points (generalization and matching via computer) are now topics 
for our current research. More precisely, an in--depth investigation of the problem will 
concern
\begin{itemize}
\item[$(i)$] extending the method to the case of different pendula, by examining equation (\ref{polcarattr}) via the spectrum analysis performed in the simpler case, or by
elaborating a formula similar to (\ref{disturb}) for evaluating the 
effects of a variance in the features of the two pendula,
\item[$(ii)$] estimating the time of decay of the motion and discard the situations where a very short time from the starting time to the almost rest state would produce not interesting cases, 
\item[$(iii)$] verifying the analytical outcome by means of simulations via computer, 
either for calculating the spectrum and for tracing the profiles of $\delta$ and $\sigma$, 
\item[$(iv)$] locating the eigenvalues in more restricted regions, by making use of some generalization of the E--K Theorem (as in \cite{gov}) which confines the spectrum in specific circular sectors of the complex plane, 
\item[$(v)$] checking whether the decrease of $\beta_i$, $i=1,2,3$ to zero in (\ref{fnlinnof}) will lead to the solutions described in Section $3$, 
\item[$(vi)$] adding the effects (\ref{escapement}) of an escapement, 
\end{itemize}

\noindent
As for the point $(i)$, the difficulty comes from the non--possibility of factorizing the characteristic equation (\ref{polcarattr}) as in (\ref{polcarip}), so that the eigenvalues for $\delta$ cannot longer be separated from the rest of the spectrum. At this point, the small coefficients ($\ell_1-\ell_2$, $\dots$) which join the equation for $\delta$ (last equation in (\ref{fnlinnof})) with $x$ and $\sigma$ will play a significant role.

\noindent
On the other hand, point $(vi)$ renders the analytical problem much complex
and deeply different from the present one: as we already remarked, the new formulation requires a non--trivial discussion of equilibrium and stability
and of the existence and the regularity of solutions, where the difficulty arises from the 
typical (but experimentally adequate) discontinuous profile of (\ref{escapement}).


\begin{thebibliography}{10}

%
\bibitem{ben} Bennett, M.~, Schatz, M.~F.~, Rockwood, H.~, Wiesenfeld, K.~, Huygens's cloks, 
Proc.~R.~Soc.~Lond.~A, {\bf 458}, 563--579 (2002)

%
\bibitem{czo} 
Czolczynki, K.~, Perlikowski, P.~, Stefanski, A.~, Kapitaniak, T.~, 
Huygens' odd sympathy experiment revisited, Int~J.~Bifurcation Chaos {\bf 21}, 2047 (2011)

%
\bibitem{dil} Dil\~ao, R.~, Anti--phase and in--phase synchronization of nonlinear oscillators: The Huygens's clocks system, Chaos {\bf 19}, 023118 (2009)

%
\bibitem{ene}
Enestr\"om, G.~, Remarque sur un th\'eor\`eme relatif aux racines de l’equation
$a_nx^n+a_{n−1}x^{n−1}+\dots a_1 x+a_0=0$ o\`u tous les coefficientes
$a$ sont r\'eels et positifs, T\^ohoku Mathematical Journal {\bf 18}, 34--36 (1920)

%
\bibitem{fra} Fradkov, A.~L.~, Andrievsky, B.~, Synchronization and phase relations in the motion of two--pendulums system, Int.~J.~of Non--Linear Mechanics, {\bf 42}, 895--901 (2007)

%
\bibitem{gan} Gantmacher, F.~.R.~, Applications of the theory of matrices,  translated and revised by J.~L.~Brenner, Interscience Publishers, Inc.~, New York   (1959)

%
\bibitem{gel} Gelfand, I.~M.~, Kapranov, M.~M.~, Zelevinsky, A.~V.~, Discriminants, resultants and multidimensional determinants, Bulletin (New Series) of the American Mathematical society {\bf 37} 2, 183--198 (1999)


%
\bibitem{gov} Govil, N.~, K.~and Rahman, Q.~I.~, On the Enestr\"om--Kakeya Theorem, 
T\^ohoku Mathematical Journal {\bf 20}, 126--136 (1968)

%
\bibitem{kak}
Kakeya, S.~, On the Limits of the Roots of an Algebraic Equation with
Positive Coefficients, T\^ohoku Mathematical Journal (First Series) {\bf 2}, 140--142 (1912--13)

%
\bibitem{kum} Kumon, M.~, Washizaki, R.~, Sato, J.~, Kohzawa, R.~, Mizumoto, I.~, Iwai, Z.~, 
Controlled synchronization of two $1$--DOF coupled oscillators, Proceedings of the $15$--th Triennal World Congress of IFAC, Barcelon, Spain (2002)

%
\bibitem{oud}
Oud, W.~, Nijmeijer, H.~, Pogromsky, A.~, Experimental results on Huygens synchronization, Proceedings of First
IFAC Conference on Analysis and Control of
Chaotic Systems, Reims,  France (2006)



\end{thebibliography}
\end{document}